\documentclass[aps,twocolumn,prd,showpacs,nofootinbib]{revtex4}
\usepackage{amsmath}
\usepackage{graphicx}
\usepackage{dcolumn}
\usepackage{bm}
\usepackage{amssymb}
\usepackage{latexsym}

\def\be{\begin{equation}}
\def\ee{\end{equation}}
\def\ba{\begin{eqnarray}}
\def\ea{\end{eqnarray}}

\begin{document}

\title{Nearly Scale-Invariant Spectrum of Adiabatic Fluctuations
May be from  a Very Slowly Expanding Phase of the Universe}

\author{Yun-Song Piao}
\email{yspiao@itp.ac.cn} \affiliation{Interdisciplinary Center of
Theoretical Studies, Chinese Academy of Sciences, P.O. Box 2735,
Beijing 100080, China}

\author{E Zhou}
\affiliation{Institute of Process Engineering, Chinese Academy of
Science, P.O. Box 353, Beijing 100080, China }

\date{August 14$^\mathrm{th}$, 2003 }

\begin{abstract}
In this paper we construct an expanding phase with phantom
 matter, in which the scale factor expands very slowly but the
 Hubble parameter increases gradually,
 and assume that this expanding
 phase could be matched to our late observational cosmology by
the proper mechanism.
 We obtain the nearly scale-invariant spectrum of adiabatic fluctuations in
 this scenario, different from the simplest inflation and usual ekpyrotic/cyclic scenario,
 the tilt of nearly scale-invariant spectrum in this scenario is blue.
 Although there exists an uncertainty surrounding the way in which
 the perturbations propagate through the
transition in our scenario, which is dependent on the detail of
possible "bounce" physics, compared with inflation and
ekpyrotic/cyclic scenario, our work may provide another feasible
cosmological scenario generating the nearly scale-invariant
perturbation spectrum.
\end{abstract}

\maketitle

The inflation scenario \cite{GL} plays an important role in modern
cosmology, which solves many problems of standard cosmology. The
physics of usual inflation models is dependent on the inflaton
potential, in general, such a potential that yields enough
e-folding number and the correct magnitude of density perturbation
is fine-tuning. A lot of observations, specifically the recent
WMAP results \cite{BHK} such as the flatness of space, the nearly
scale-invariance, adiabaticity and Gaussian of the density
perturbations imply that the inflation is very consistent early
cosmological scenario. But the inflation may be not uniquely
consistent scenario with the WMAP data. There remains some
alternatives \cite{BLOS}. An example is the ekpyrotic/cyclic
scenario \cite{KOS, TTS}, which is motivated by the string/M
theory, in which the visible universe is a boundary brane in five
dimensional bulk space-time and the collision between two boundary
branes leads to the reheating in visible universe corresponding to
the Big Bang of standard cosmology. The relevant dynamics can be
described by a 4D effective theory in which the separation of the
branes in the extra dimensions is modeled as a scalar field. In
this scenario, the perturbation leaving the horizon during the
contracting phase reenters the horizon after the bounce to an
expanding phase corresponding to our observational cosmology. If
the proper matching conditions during the bounce is considered
\cite{TTS, D, TT}, see also \cite{PP, CDC}, the nearly
scale-invariant spectrum could be obtained.

Both the inflation and the ekpyrotic/cyclic models rely on the
parameter $\omega$ of state equation having a specific qualitative
behavior throughout the period when the perturbations are
generated. For inflation, the condition on $\omega$ is $
\omega\simeq -1$ and for ekpyrotic/cyclic model, it is $\omega\gg
1$. Correspondingly, the Hubble parameter is nearly constant
during inflation, and the 4D scale factor contracts very slowly
and is nearly constant during ekpyrosis/cyclic. In some sense,
both scenarios can give some results satisfying the WMAP
observation. The spectrum is nearly scale-invariant and the tilt
of spectrum is red in their simplest realization \cite{GKS}. In
this paper, an early universe model with the phantom matter
\footnote{The state equation of matter is $p\equiv \omega \rho$.
The matter with the state parameter $\omega <-1$, dubbed "phantom
matter" in general, has received increased attention recently
\cite{CKW, CHT}, especially applied to the late universe as
explanation for the dark energy, because it has some strange
properties. The phantom matter violates the dominant-energy
condition prohibiting time machines and wormholes, and its energy
density increase with time, and may be up to infinite in a finite
time and lead to "Big Rip". But in our work, we focus on the early
universe, especially the generation of nearly scale-invariant
adiabatic spectrum. } is proposed, in which the 4D scale factor
expands very slowly and is nearly constant which corresponds to
$\omega \ll -1$. We find that when this expanding phase is
"linked" to another expanding phase with usual radiation and
matter by the proper mechanism, a nearly scale-invariant spectrum
may be generated and the tilt of spectrum is blue.

We start with such a 4D effective action of phantom field as
follows \be {\cal L}_{\rm mat}= {1\over
2}(\partial_{\mu}\varphi)^2 - V(\varphi) \label{mat} \ee where the
metric signature $(-~+~+~+)$ is used, If taking the field
$\varphi$ spatially homogeneous but time-dependent, the energy
density $\rho$ and pressure $p$ can be written as \be \rho
=-{1\over 2}{\dot \varphi}^2 + V(\varphi) ~~~~~ p=-{1\over 2}{\dot
\varphi}^2-V(\varphi) \label{p} \ee From (\ref{p}), for $\rho
>0$ the state parameter $\omega \equiv {p\over \rho}< -1$ can be
seen. We minimally couple the action (\ref{mat}) to the 4D
gravitational action \be {\cal L}_{\rm gra} = {1\over 16\pi G} R
\label{gra} \ee where $G$ is the Newton gravitational constant,
$R$ is the 4D curvature scalar. The Friedmann universe, described
by the scale factor $a(t)$, satisfies the equations \be h^2 =
{8\pi G\over 3} \left(-{1\over 2 }{\dot \varphi}^2
+V(\varphi)\right) \label{da} \ee and the dynamical equation of
phantom field is \be {\ddot \varphi} +3h{\dot \varphi} -
V^\prime(\varphi)=0 \label{drho} \ee where $h ={{\dot a}\over a}$
is the Hubble parameter. Combining (\ref{da}) and (\ref{drho}),
the relation \be \epsilon\equiv {3\over 2}(1+\omega) = -{{\dot
h}\over h^2} \label{ome}\ee can be obtained. Since $\omega(t)<
-1$, considering (\ref{ome}), ${\dot h} >0 $ is required.
Therefore, a reasonable selection for the scale factor $a(t)$ is
\be a(t) \sim {1\over (-t)^{n(t)}}\label{at} \ee For $t$ initially
from $-\infty$ to $0_-$, it corresponds to an expanding phase. We
assume ${-t\over n}{dn\over dt} \ll 1$, {\it i.e.} the variable
rate of $n(t)$ is very small and near constant, and have \be
h\simeq {n\over (-t)} ~~~~~ {\dot h}\simeq {n\over (-t)^2}
\label{h} \ee thus \be\epsilon \simeq -{1\over n}\label{eps}\ee
Consequently, the background values of all relevant quantities can
be determined. From (\ref{da}), (\ref{drho}) and (\ref{h}), \be
{\dot \varphi}^2 \simeq {n\over 4\pi G} {1\over
(-t)^2}\label{dotphi}\ee can be given, thus \be V(\varphi) \simeq
{n(3n +1) \over 8\pi G}{1\over (-t)^2} \label{vphi}\ee We see that
for arbitrary value $n$, $V(\varphi)$ is positive, which is
different from the usual scalar field, in which $n<{1\over 3}$,
the potential is negative. From (\ref{dotphi}) and (\ref{vphi}),
we can also check and obtain (\ref{eps}). Integrating
(\ref{dotphi}), and instituting this result into (\ref{vphi}), we
derive the effective potential \be V(\varphi)\simeq {n(3n+1)\over
8\pi G} \exp{\left(-\sqrt{16\pi G\over n}
\varphi\right)}\label{vv}\ee

Instituting (\ref{dotphi}) and (\ref{vphi}) into (\ref{p}), we see
that the energy density of phantom field \be \rho\simeq {3n^2\over
8\pi G}{1\over (-t)^2} \sim {3n^2\over 8\pi G} a^{2\over n}
\label{phan}\ee increases with expansion, while the energy density
of usual matter and radiation decreases respectively at \be
\rho_{\rm matt} \sim {1\over a^3}~~~~~\rho_{\rm rad}\sim {1\over
a^4}\label{mr}\ee Thus the evolving solution dominated by phantom
field will be an attractor, which is generic and likes the case
that the phantom matter is regarded as the dark energy.

When $t \rightarrow 0_{-}$, the Hubble parameter $h$ increases
gradually, a singularity will appear. For the transition to the
observational cosmology, we may expect that at some time the
phantom field could decay into the usual radiation by some
coupling \footnote{The phantom matter may be unstable and will
decay into the ordinary particles and other phantom particles,
whose lifetimes depends on the cutoff scale \cite{CHT}, which in
some sense, is similar to the reheating mechanism after usual
inflation.
} or through the singular "Big Rip" by some other mechanism from
high energy/dimension theory, which may be regarded as a "bounce"
to an observational cosmology, like Pre-Big-Bang \cite{GV} (see
\cite{V} for recent reviews) and ekpyrotic/cyclic scenario.


We discuss the metric perturbations of this scenario in the
following. In longitudinal gauge and in absence of anisotropic
stresses, the scalar metric perturbation can be written as \be
ds^2 = a^2(\eta) (-(1+2\Phi)d\eta^2 +(1-2\Phi) \delta_{ij}dx^i
dx^j )\ee where $\Phi$ is the Bardeen potential and $\eta$ is
conformal time $ d\eta ={dt\over a}$, thus \be \eta\sim
-(-t)^{n+1} \label{eta} \ee \be a(\eta) \sim (-\eta)^{-{n\over
n+1}} \simeq (-\eta)^{1\over \epsilon -1}\label{aeta}\ee \be
{\tilde h} ={a^\prime \over a^2}\simeq {1\over (\epsilon -1) a
\eta} \label{hti}\ee In general the curvature perturbation $\zeta$
on uniform comoving hypersurface \be \zeta = -{h\over {\dot
h}}\left({\dot \Phi} +h\Phi\right) +\Phi \label{zeta}\ee and the
Bardeen potential $\Phi$ change dramatically across the bounce,
thus the usual matching conditions used in inflation model may be
not proper. More recently, as is pointed to in Ref. \cite{CDC},
the resulting spectral index in late radiation-dominated universe
depends on which of $\zeta$ and $\Phi$ passes regularly through
the bounce, which is dependent on detailed features of the
scenarios at the "bounce". At the moment we can hardly describe
the detail of "bounce" physics, and paying attention to the
generation of nearly scale-invariant spectrum of adiabatic
fluctuations, we assume that the Bardeen potential $\Phi$ is
regular through the "bounce" from the pre-expanding phase with
phantom field to late expanding phase with usual radiation.
and straightly focus on the evolution of the Bardeen potential
$\Phi$ \footnote{In fact, due to the lack of "bounce" physics,
there may exist an uncertainty about the perturbation spectrum,
which is from the possible effects of "bounce" on $\zeta$ and
$\Phi$ propagating through the transition. We will go back to this
issue in the future. Furthermore, we may connect this scenario
with recent observation results, such as the low quadrupole and
running index, in our coming works. } .

We define $u\equiv {a \Phi\over \varphi^\prime}$ and obtain the
differential equation that the Fourier mode $u_k$ of $u$ obeys \be
u_k^{\prime\prime} +\left(k^2 - {\beta(\eta)\over \eta^2}\right)
u_k =0 \label{uk}\ee with \ba &&\beta(\eta)\equiv  \eta^2 {\tilde
h}^2 a^2 \{\epsilon -{(1-\epsilon^2)\over
2}\left({d\ln{|\epsilon|}\over d{\cal N}}\right) \nonumber \\ & +&
 {(1-\epsilon^2)\over 4}\left({d\ln{|\epsilon|}\over d{\cal
N}}\right)^2 -{(1-\epsilon)^2\over
2}\left({d^2\ln{|\epsilon|}\over d{\cal N}^2}\right)\}\ea in which
(\ref{at}) and (\ref{eta}) are considered, the prime denotes
differentiation with respect to the conformal time $\eta$ and the
variable $\cal N$, which measures the e-folding number which exits
the horizon before the end of the pre-expanding phase, \be {\cal
N} \equiv \ln{\left({a_e {\tilde h}_e\over a {\tilde
h}}\right)}\ee where the subscript 'e' denotes the quantity
evaluated at the end of the expanding phase. Assuming that $\beta$
is near constant for all interesting modes $k$, we can solve
(\ref{uk}) analytically and obtain \be u_k = \sqrt{-k\eta}\left(
C_1(k)J_v(-k\eta)+C_2(k)J_{-v}(-k\eta)\right)\label{uks}\ee where
$v\equiv \sqrt{\beta +{1\over 4}}$ \footnote{For $a\sim
(-\eta)^q$, $q$ is constant, $\beta = q(q+1)$, thus $\beta
+{1\over 4}= (q+{1\over 2})^2>0$ is always satisfied, independent
the value of $q$. In our model, $q\simeq {-1\over 1-\epsilon}$. },
and $J_v$ is the first kind of the Bessel function with order $v$
and the function $C_i(k)$ can be determined by specifying the
initial conditions.

During the pre-expanding, the Hubble parameter $h$ increases
gradually, thus the initial perturbation in the horizon will exit
the horizon, and reenter the horizon after the "bounce" to an
expanding phase corresponding to our observational cosmology in
which the Hubble parameter $h$ decreases gradually. In the regime
$k^2\eta^2 \gg |\beta|$, in which the modes $u_k$ is very deep in
the horizon, the mode equation (\ref{uk}) reduced to the equation
for a simple harmonic oscillator and $u_k$ is stable. In this
limit, we may make the usual assumption that at $\eta\rightarrow
-\infty$ the normalized initial conditions is \be u_k \sim
{e^{-ik(\eta-\eta_i)} \over (2k)^{3\over 2}}\ee where $\eta_i$ is
an arbitrary conformal time having no influence on the subsequent
evolution. In the regime $k^2\eta^2 \ll |\beta|$, in which the
mode $u_k$ is far out the horizon, the mode is unstable and grows.
In long-wave limit, $\Phi_k$ can be given and expanded to the
leading term of $k$ \be k^{3\over 2}\Phi_k \sim k^{-v+{1\over
2}}\simeq k^{-\beta} \ee Thus the corresponding spectrum index can
be regarded as \be n_s -1 \simeq -2\beta \simeq -{2\over
(1-\epsilon)^2}\left\{\epsilon -{(1-\epsilon^2)\over
2}\left({d\ln{|\epsilon|}\over d{\cal
N}}\right)\right\}\label{ns}\ee where in the second equation, the
higher order terms $\left({d\ln{|\epsilon|}\over d{\cal
N}}\right)^2$ and ${d^2\ln{|\epsilon|}\over d{\cal N}^2}$ have
been neglected. We see, from (\ref{ns}), that since $\epsilon <0$,
the nearly scale-invariant spectrum requires $\epsilon \ll 0$
\footnote{In fact, when $\epsilon \simeq 0$, the nearly
scale-invariant spectrum can be also obtained, which corresponds
to $\omega\simeq -1$, and may be regarded as the phantom
inflation.}, which corresponds to $\omega \ll -1$, {\it i.e.} \be
\omega +1 ={\rho+p \over \rho} ={-{\dot \varphi}^2\over -{1\over
2}{\dot \varphi}^2 +V(\varphi)}\ll 0 \ee Considering (\ref{eps}),
$n\sim 0$ is required, {\it i.e.} the expansion of the scale
factor is very slow, thus from (\ref{vv}), we see that the
effective potential of phantom field is very steep and the phantom
field rolls up quickly during a period of very slow expansion. In
this case, we have \be n_s-1 \simeq {2\over |\epsilon |}-
{d\ln{|\epsilon|}\over d{\cal N}}\ee Since $a(\eta)$ is near
constant, the end of this very slowly expanding phase may occur
when $a(\eta)$ begins to change significantly. Similar to the
analysis of Ref. \cite{GKS}, from (\ref{aeta}) and (\ref{hti}), we
obtain \be a\sim \left({1\over (1-\epsilon) a{\tilde
h}}\right)^{1\over \epsilon -1}\ee thus for $\epsilon \simeq
const$, we have \be {a_e\over a}=\left({a{\tilde h}\over
a_e{\tilde h}_e}\right)^{1\over \epsilon -1}\simeq e^{-{{\cal
N}\over \epsilon}}\ee Since in such a very slowly expanding phase,
the significant changing of $a$ can be written as ${a_e\over a}
\simeq e$, one obtains \be \epsilon \simeq -{\cal
N}\label{epsi}\ee Thus combining (\ref{eps}) and (\ref{epsi}), we
have \be n_s-1 \simeq {2\over {\cal N}} -{1\over {\cal N}} =
{1\over {\cal N}}\simeq n\ee Different from the usual
ekpyrotic/cyclic models, since $n$ is very small, a nearly
scale-invariant blue spectrum is obtain.

In summary, we construct an expanding phase with the phantom
matter, in which the scale factor expands very slowly. We expect
that at some time the phantom matter may decay into the usual
radiation by some coupling, or through the singular "Big Rip" into
the radiation-dominated phase by some other mechanism, which may
be from the string/M theory or other high energy/dimension theory.
If regarding such a transition as a "bounce" from this
pre-expanding phase to the late expanding phase denoting our
observational cosmology, after the "bounce", the nearly
scale-invariant spectrum of adiabatic fluctuations may be obtain.
Different from ekpyrotic/cyclic scenario in which the bounce is
from the contracting phase to the expanding phase, the "bouncing"
in our scenario is from the expanding phase to the expanding
phase, however, instead of the red spectrum in its simplest
realization, the spectrum is blue in our scenario. We also gave a
phantom field action describing such a very slowly expanding
phase. Since in our scenario, the phantom field will roll up
quickly along its steep effective potential, after it decay or
pass through the singularity, it may be placed again in the bottom
of the effective potential, then roll up again, which may lead to
a variable "cyclic" scenario. Compared with usual inflation and
ekpyrotic/cyclic scenario, our work may provide another feasible
cosmological scenario to generate the nearly scale-invariant
perturbation spectrum, which is worth studying further.




\begin{thebibliography}{99}


\bibitem{GL} A.H. Guth, Phys. Rev. \textbf{D23} (1981) 347;
A.D. Linde, Phys. Lett. \textbf{B108} (1982) 389; A.A. Albrecht
and P.J. Steinhardt, Phys. Rev. Lett. \textbf{48} (1982) 1220.

\bibitem{BHK} C. Bennett et al., astro-ph/0302207;
G. Hinshaw et. al., astro-ph/0302217; A. Kogut et. al.,
astro-ph/0302213.

\bibitem{BLOS} S.L. Bridle, O. Lahav, J.P. Ostriker and P.J.
Steinhardt, Science \textbf{299} (2003) 1532, astro-ph/0303180.

\bibitem{KOS} J. Khoury, B.A. Ovrut, P.J. Steinhardt and N. Turok,
Phys. Rev. \textbf{D64} (2001) 123522; J. Khoury, B.A. Ovrut, N.
Seiberg, P.J. Steinhardt and N. Turok, Phys. Rev. \textbf{D65}
(2002) 086007; P.J. Steinhardt and N. Turok, Science \textbf{296},
(2002) 1436; Phys. Rev. \textbf{D65} 126003 (2002).

\bibitem{TTS} A.J. Tolley, N. Turok and P.J. Steinhardt,
hep-th/0306109.

\bibitem{D} R. Durrer, hep-th/0112026; R. Durrer and F. Vernizzi,
Phys. Rev. \textbf{D66} (2002) 083503.

\bibitem{TT} A. Tolley and N. Turok, Phys. Rev. \textbf{D66}
(2002) 106005, hep-th/0204091.

\bibitem{PP} P. Peter and N. Pinto-Neto, hep-th/0203013; P.
Peter, N. Pinto-Neto and D.A. Gonzalez, hep-th/0306005.

\bibitem{CDC} C. Cartier, R. Durrer and E.J. Copeland,\
hep-th/0301198.

\bibitem{GKS} S. Gratton, J. Khoury, P.J. Steinhardt and N. Turok,
astro-ph/0301395; J. Khoury, P.J. Steinhardt and N. Turok,
astro-ph/0302012.

\bibitem{CKW} G.W. Gibbons, hep-th/0302199;
R.R. Caldwell, M. Kamionkowski and N.N. Weinberg,
astro-ph/0302506; S. Nojiri and S.D. Odintsov, hep-th/0303117;
hep-th/0304131; hep-th/0306212; P. Singh, M. Sami and N. Dadhich,
hep-th/0305110; A. Yurov, astro-ph/0305019; P.F. Gonzalez-Diaz,
astro-ph/0305559.

\bibitem{CHT} S.M. Carroll, M. Hoffman and M. Trodden, astro-ph/0301273



\bibitem{GV} M. Gasperini and G. Veneziano, Astropart. Phys.
\textbf{1} (1993) 317, hep-th/9211021.

\bibitem{V} G. Veneziano, hep-th/0002094; J.E. Lidsey, D. Wands
and E.J. Copeland, Phys. Rept. \textbf{337} (2000) 343,
hep-th/9909061; M. Gasperini and G. Veneziano, Phys. Rept.
\textbf{373} (2003) 1, hep-th/0207130.



\end{thebibliography}
\end{document}